# Longitudinal Schottky spectra of a bunched Ne$^{10+}$ ion beam at the CSRe

WEN Wei-Qiang (汶伟强)[1,2]   MA Xin-Wen (马新文)[1]   ZHANG Da-Cheng (张大成)[1]
ZHU Xiao-Long (朱小龙)[1]   MENG Ling-Jie (孟令杰)[1]   LI Jie (李杰)[1,2]   LIU Hui-Ping (刘惠萍)[1]
ZHAO Dong-Mei (赵冬梅)[1]   WANG Zhi-Shuai (王之帅)[1,2]   MAO Rui-Shi (毛瑞士)[1]
ZHAO Tie-Cheng (赵铁成)[1]   WU Jun-Xia (武军霞)[1]   MA Xiao-Ming (马晓明)[1]   YAN Tai-Lai (晏太来)[1]
LI Guo-Hong (李国宏)[1]   YANG Xiao-Dong (杨晓东)[1]   LIU Yong (刘勇)[1]   YANG Jian-Cheng (杨建成)[1]
YUAN You-Jin (原有进)[1]   XIA Jia-Wen (夏佳文)[1]   XU Hu-Shan (徐瑚珊)[1]   XIAO Guo-Qing (肖国青)[1]
ZHAO Hong-Wei (赵红卫)[1]

[1]Institute of Modern Physics, Chinese Academy of Sciences, Lanzhou 730000, China
[2]University of Chinese Academy of Sciences, Beijing 100049, China

**Abstract:** The longitudinal Schottky spectra of a radio-frequency (RF) bunched and electron cooled $^{22}$Ne$^{10+}$ ion beam at 70 MeV/u have been studied by a newly installed resonant Schottky pick-up at the experimental cooler storage ring (CSRe), at IMP. For an RF-bunched ion beam, a longitudinal momentum spread of $\Delta p/p = 1.6 \times 10^{-5}$ has been reached with less than $10^7$ stored ions. The reduction of momentum spread compared with coasting ion beam was observed from Schottky noise signal of the bunched ion beam. In order to prepare the future laser cooling experiment at the CSRe, the RF-bunching power was modulated at 25$^{th}$, 50$^{th}$ and 75$^{th}$ harmonic of the revolution frequency, effective bunching amplitudes were extracted from the Schottky spectrum analysis. Applications of Schottky noise for measuring beam lifetime with ultra-low intensity of ion beams are presented, and it is relevant to upcoming experiments on laser cooling of relativistic heavy ion beams and nuclear physics at the CSRe.

## 1. Introduction

Schottky diagnostic has been applied in many storage rings to diagnose the revolution freqeuncy and momentum spread of the ion beams [1, 2]. As a main diagnostic tool at storage rings, Schottky diagnostic has many advantages, e.g. non-perturbing and high sensitivity. It has been employed to monitor the cooling processes, such as the stochastic cooling, the electron cooling and the laser cooling with coasting ion beams, as well as bunched ion beams [3-5]. The Schottky spectrum of low intensity heavy ion beams was also applied to mass measurement of radioactive nuclear at the storage ring, which has already been successfully performed at the experimental storage ring ESR [6], and will be applied at the CSRe as well. In case of extremely well-cooled beams featuring ordering effects [4, 7] the relative momentum spread can be as low as $\Delta p/p = 5.0 \times 10^{-7}$, the Schottky diagnostic system provides a unique method to monitor this phase transition and the dynamics of the ion beams from a gaseous state to an ordering state, even to a crystalline beam [8-11].

In this paper, the longitudinal Schottky spectra of the RF-bunched and electron cooled $^{22}$Ne$^{10+}$ ion beam at an energy of 70 MeV/u measured by a newly installed resonant Schottky pick up are presented. Applications of the Schottky noise spectra of bunched ion beams are presented and discussed, e.g. lifetime measurement at ultra-low intensity and precision atomic physics experiments at the CSRe.

## 2. Experimental setup

The experiment was carried out at the CSRe at IMP, Lanzhou. The schematic view of the CSRe is shown in Fig. 1. An electron cooler system [12] is installed to provide ion beams of high quality for performing the precision nuclear and atomic experiments [13, 14]. In order to diagnose ion beam parameters with high resolution and perform the nuclear mass measurement, a new resonant Schottky pick-up [1] was constructed and installed at one straight section of the CSRe [15]. For the purpose of performing the upcoming laser cooling of relativistic ion beams at the CSRe, a radio-frequency bunching system (RF-buncher) was installed to bunch ion beams and to provide a longitudinal force that counteracts the laser scattering force [16].

An ion beam of $^{22}$Ne$^{10+}$ was provided by an Electron Cyclotron Resonance (ECR) ion source [17] and accelerated by a Sector Focus Cyclotron (SFC), and then ion beams were injected into the main Cooler Storage Ring (CSRm). After acceleration at the CSRm, ion beams were extracted, and injected into the CSRe at an energy of 70MeV/u corresponding to an ion velocity of 37% of the speed of light. The circumference of the CSRe is 128.8 m, the revolution frequency at the chosen beam energy is 0.85 MHz. For each injection, the RF-buncher was operated at various harmonic (h=25, 50, 75) of the revolution frequency to capture and bunch the ion beam, the RF-bunching frequency is $f_{bunch} = hf_{revolution} = h \cdot \upsilon / C$, $\upsilon$ and $C$ are the velocity of the ion beam and the circumference of the storage ring, respectively. The RF-bunched ions inside of the CSRe will stay and oscillate in the bucket pseudo-potential which is produced by the RF-buncher. In order to reduce momentum spread of the ion beam, the electron cooler was operated continously during the experiment. The ion beam current was measured by a DC current transformer (DCCT). The revolution frequency and the longitudinal momentum spread of the bunched ion beam were measured by the newly installed resonant Schottky pick-up, which has much higher sensitivity than the convertional Schottky pick-up device [1]. The parameters of the bunched ion beam such as synchrotron frequency and effective RF-bunching amplitude could be extracted from Schottky spectra. This new resonant Schottky pick-up system will be also used to diagnose the state of the ion beam during the upcoming experiment of laser cooling of relativistic ion beams at the CSRe, especially for the ultra-low intensity and ultra-cold ion beams.

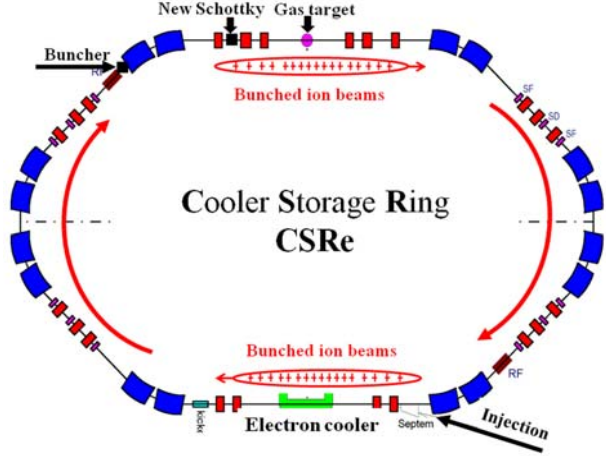

Fig. 1. (color online) Schematic view of the heavy ion storage ring CSRe at IMP. The locations of the electron cooler, new Schottky pick-up and RF-buncher are shown.

## 3. Schottky spectra of bunched ion beams

In this experiment at the CSRe, only longitudinal Schottky signals were measured since the resonant Schottky pick-up is not able to measure the transverse signals. The random signals induced by ions circulating in the storage ring will be detected by the Schottky pick-up. These signals were recorded by a spectrum analyzer, and the frequency spread of the ion beam can be obtained by Fast Fourier Transformer (FFT) at the chosen harmonic number. The relative revolution frequency spread of the Schottky signal is proportional to the relative momentum spread of the ion beam, which at the 1$^{st}$ harmonic of the revolution frequency can be written as

$$\frac{\Delta p}{p_0} = \frac{1}{\eta} \frac{\Delta f}{f_0} \quad (1)$$

where $f_0$ is the revolution frequency, $\eta = 1/\gamma^2 - 1/\gamma_t^2$ [7] is the frequency dispersion, $\gamma$ ($\gamma = 1/\sqrt{1+\beta^2}$) is the relativistic Lorentz factor, and $\gamma_t$ is the transition factor of the machine.

Since the resonant Schottky pick up works at the resonant frequency of about 243 MHz and the revolution frequency of the ion beam is $f_0 = 0.85$ MHz, the 285$^{th}$ harmonic number was selected to match the resonant frequency. A typical Schottky spectrum of RF-bunched ion beam at the 50$^{th}$ harmonic of the revolution frequency is shown in Fig. 2. The center frequency of the Schottky spectrum is 243.377813 MHz, and the span is 10 kHz. Many small peaks appear for bunched ion beams compared to the Schottky spectrum obtained from coasting ion beams [5]. This was caused by the oscillation of ions inside of the bucket which is produced by the RF-buncher. Each small peak in the spectrum represents a synchrotron satellite. A detailed description of longitudinal Schottky signal of bunched ion beam can be found in Ref. [18].

The width of the Schottky spectrum is the product of the harmonic number and the revolution frequency distribution of the particles. In Fig. 2, the black solid line is the measured result, the red solid line is the Gausian fit of the Schottky spectrum. By using Eq (1), the longitudinal momentum spread obtained is about $\Delta p/p \approx 1.6 \times 10^{-5}$ with less than $10^7$ stored ions after electron cooling.

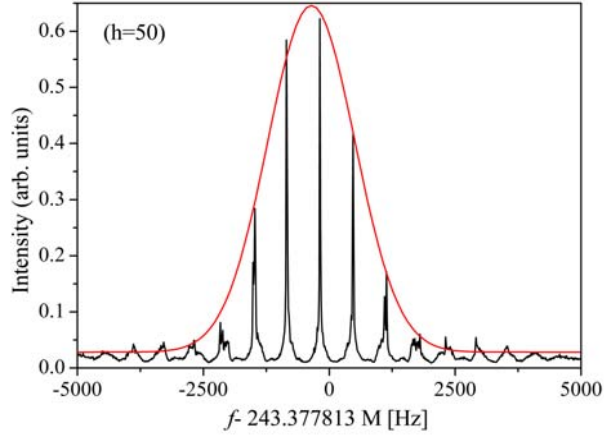

Fig. 2. Schottky spectrum of bunched $Ne^{10+}$ ion beam at the 50$^{th}$ harmonic of revolution frequency (black line ), x axis is the frequency domain and y axis is the Schottky noise power intensity. The Gaussian function was used to fit the spectrum (red solid line), and the obtained momentum spread is about $\Delta p/p \approx 1.6 \times 10^{-5}$.

## 4. Momentum spread comparison with bunched and coasting beams

The Schottky spectra of a bunched ion beam (black solid line) and a coasting ion beam (red solid line) are shown in Fig. 3. The RF-buncher was operated at 50$^{th}$ harmonic number of the revolution frequency for the bunched ion beam. It needs to be noted that, these two Schottky signals were measured with the same beam current (10 μA). The frequency distribution of the bunched ion beam is much narrower compared to the Schottky signals of coasting ion beam, corresponding to the momentum spread of the bunched ion beam is much lower than the coasting ion beam. There are two reasons need to be considered to interpret this phenomenon, on the one hand, the bucket potential produced by the RF-buncher has an acceptance of the longitudinal momentum spread which is about $\Delta p/p \approx 5.5 \times 10^{-5}$. Therefore, most of the ions were forced into the bucket, and the too hot part of the ion beam can not be captured and will be lost during the experiment. On the other hand, the RF-buncher will minimize velocity changing effects which are due to fluctuations in the voltage of the electron cooler, this has already been successfully applied at the laser spectrocopy experiment at the ESR [19]. As a result, the RF-buncher applied to coasting ion beams provided a possibility to reduce the longitudinal momentum spread, and this method could be used for the future precision atomic or nuclear experiments at the CSRe [20, 21].

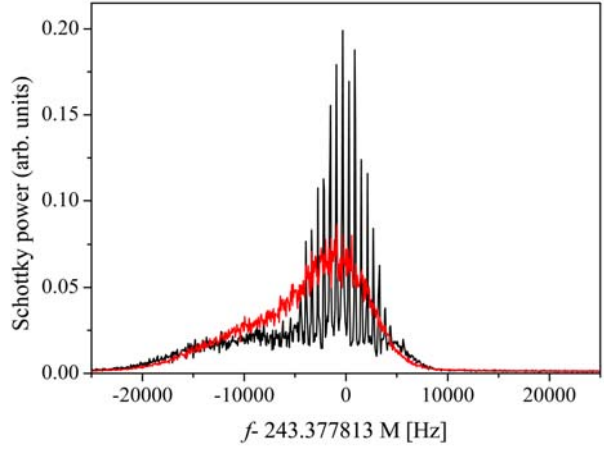

Fig. 3. (color online) Schottky signal of a RF-bunched (black solid line) and a coasting (red solid line) $^{22}Ne^{10+}$ ion beam circulating at the CSRe. The RF-buncher was operated at the 50$^{th}$ harmonic number, and both of the Schottky spectra were taken with the same beam current (10 μA) after electron cooling.

## 5. Effective RF-bunching amplitude measured by the Schottky spectrum

In the longitudinal Schottky sprectrum of a RF-bunched and electron-cooled ion beam, as shown in Fig. 2, the distance of every two adjacent small peaks is the synchrotron frequency of the ions oscillating inside the bucket. This synchrotron freqency of the ions in the bucket can be written as [4]

$$\omega_s = \frac{\omega_{rev}}{\beta}\sqrt{\frac{qeh\eta U_b}{2\pi\gamma mc^2}} \qquad (2)$$

It depends on the revolution frequency $\omega_{rev}$, the beam velocity $\beta c$, the ion charge $qe$, the harmonic number $h$, the frequency dispersion $\eta$, the effective RF bunching amplitude $U_b$, and the relativistic Lorentz factor $\gamma$, $m$ is the mass of the ion and $c$ is the speed of light. This allows us to measure the RF-bunching amplitude precisely, and it is actually the only way to measure the effective bunching amplitudes at the storage rings.

In order to produce different counteractive force to counteract with the laser scattering force in the upcoming laser cooling experiment[16], the ion beam was RF-bunched at various harmonic numbers ($h$ = 25, 50, 75), and amplitudes of the voltage (Vpp) added on the RF-buncher at each harmonic number were tuned from 500 mV to 200 mV during the experiment. The Schottky spectra were recorded at each injection. From these Schottky spectra the synchrotron freqencies of ions inside the bucket were extracted (distance between every two adjacent small peaks). By making use of the Eq. (2), effective bunching amplitudes of ions experiencing in the bucket produce by the RF-buncher at the 25$^{th}$, 50$^{th}$ and 75$^{th}$ harmonic numbers are shown in Fig. 4. The detailed disscusion of the longitudinal dynamic of the bunched ion beam during this experiment, especially the dynamics of the ions oscillate inside the bucket will be given else where. It can be seen that, the effective RF-bunching amplitudes are all higher than 5 V which is needed for the laser cooling experiment. Since the laser scattering force added on the ions during laser cooling experiments is much samller than that provided by the RF-buncher [8] here, the RF-amplitudes should be reduced by decreasing the signal generator's output signal amplitudes during the future laser cooling experiment at the CSRe.

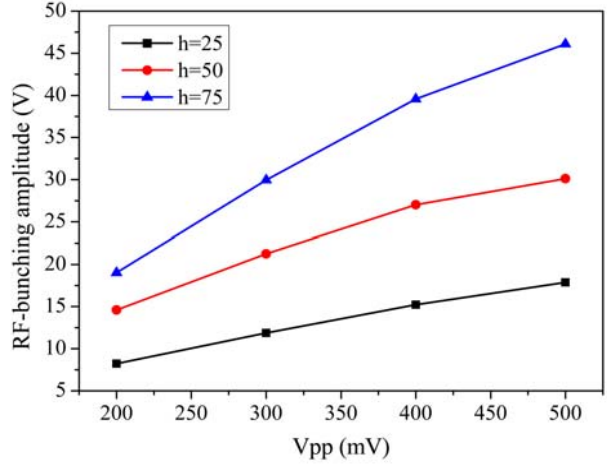

**Fig. 4.** (color online) The bunching amplitudes of ions experienced inside the bucket which are produced by the RF-buncher. These bunching amplitudes were extracted from the Schottky spectrum recorded at 25$^{th}$, 50$^{th}$ and 75$^{th}$ harmonic of the revolution frequency with different applitudes of the signals.

## 6. Lifetime measurement by the Schottky spectra

The lifetime of ion beams circulating at storage rings is commonly measured by the DCCT. Unfortunately, the DCCT has low sensitivity (1 μA at the CSRe), thereby the Schottky noise power can be utilized to measure the lifetime and ion number with ultra-low beam intensities ($N<10^3$). The power of Schottky spectrum $P(f)$ which is measured by the Schottky pick-up and following spectrum analyzer is given by [22]

$$P(f) = Z_t I_{rms}^2 = 2Z_t f_0^2 q_e^2 N \qquad (3)$$

where $Z_t$ is the transfer impedance from current to voltage, $f_0$ the revolution frequency, $q_e$ the charge of the ions, and $N$ is the number of stored ions. The noise power is constant at each harmonics and is proportional to the number of stored ions. Therefore, the Schottky noise power can be used to measure beam lifetimes at the storage ring.

By tuning the spectrum analyzer to a fixed frequency band, which is centered on the momentum spectrum and wide enough to span the momentum spread of the ion beam, one obtains a Schottky signal proportional to the beam intensity. The beam current measured by DCCT and Schottky noise power measured by resonant Schottky pick-up as a function of time are shown in Fig. 5 (a) and (b), respectively. Every data point in Fig. 5 (b) extracted from the Schottky spectrum was averaged with 20 frames (16 ms for each frame), which is about 0.36 s. On a linear scale, exponential decay curves (red line) are fitted to the experimental signals in Fig. 5, and lifetime τ = 93.5 ± 4.5 s concluded from the decrease of the Schottky noise power at low beam intensity (see Fig. 5 (b)) is in good agreement with the lifetime τ = 90.7 ± 2.3 s measured by the DCCT (see Fig. 5 (a)). However, the beam lifetime measured by Schottky noise signal has lower precision than the measurement by the DCCT in this experiment, the optimum of the Schottky signal is still needed in the further experiments.

Since the DCCT installed at the CSRe has a sensitivity of 1μA ($N = 10^7$ in this experiment), therefore, it will not be sensitive if less number of ions stored at the storage ring. Therefore, as discussed in previous part, the integrated Schottky noise power can be used to determine the number of stored particles even for ultra-low intensity ($N < 10^3$). Fortunately, the newly installed resonant Schottky pick-up at the CSRe has an ultra-high sensitivity and a possibility to take valuable spectra in short time [1]. It thus will be a powerful tool for diagnosing low beam intensity of ultra-cold ion beams, especially for measuring the nuclear mass with electron cooling [23].

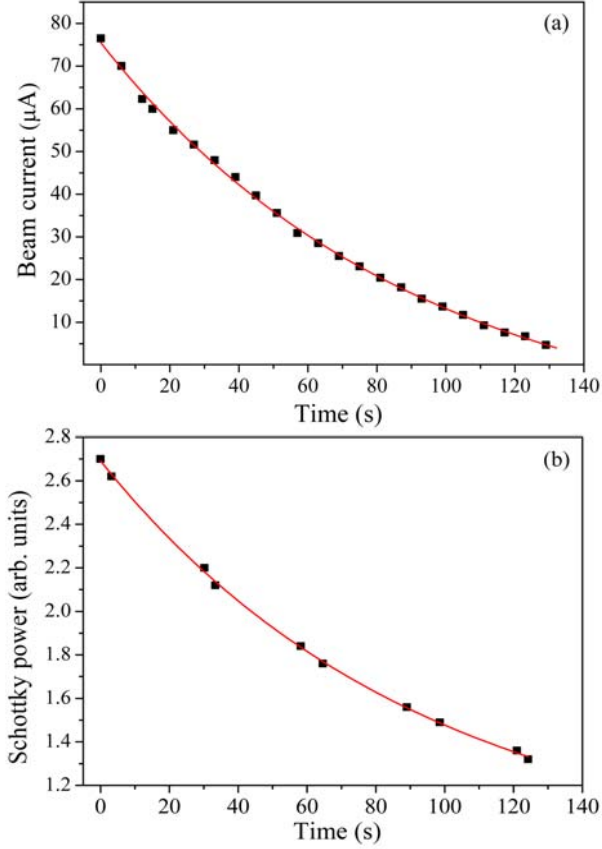

Fig. 5. Beam current measured by the DCCT (a) and Schottky noise power measured by the resonant Schottky pick-up (b) as a function of time. The exponential function was used to fit (red solid line) the measured data and obtained lifetimes of the ion beam measured by DCCT and Schottky noise power are 90.7 ± 2.3 s and 93.5 ± 4.5 s, respectively.

## 7. Conclusion

The longitudinal Schottky spectra of an RF-bunched and electron-cooled ion beam were measured by a newly installed resonant Schottky pick-up at the CSRe. The experiment was carried out with a $^{22}Ne^{10+}$ ion beam at an energy of 70 MeV/u. Compared to the coasting ion beam, the reduction of relative momentum spread of the bunched ion beam was observed. For the purpose of performing laser cooling experiment at the CSRe, the newly installed RF-buncher was operated at various harmonic of revolution frequency. From the synchrotron frequencies of the ions oscillating inside of the bucket, the effective RF-bunching amplitudes at various harmonic of the revolution frequency were extracted, which could be very helpful for the upcoming laser cooling experiment. The lifetime of ion beam measured by the Schottky noise at ultra-low intensitis is in good agreement with the result from the DCCT, which are strongly relevant to the future experiments with ultra-cold ion beams, e.g. radioactive nuclear mass measurement [20] and laser cooling of relativistic heavy ion beams.

## Acknowledgements

WEN Wei-Qiang would like to thank Dr. Michael Bussmann for his help and valuable discussions. The authors would like to thank the crew of Accelerator Department for their skillful operation of the CSR accelerator complex. This work has been supported by the National Natural Science Foundation of China (10921504).